\documentclass[twocolumn,letterpaper,superscriptaddress,showpacs,prb]{revtex4}
\usepackage{graphicx}
\usepackage[usenames]{color}
\usepackage{amsmath}
\usepackage{amssymb}
\usepackage{mathrsfs}
\usepackage{url}

\newcommand{\tr}{\mathop{\mathrm{Tr}}}
\newcommand{\Span}{\mathop{\mathrm{Span}}}

\begin{document}
% versione meilleure jusqu'a present :
\title{Quantum transport through resistive nanocontacts:\\
Effective one-dimensional theory and conductance formulas for 
non-ballistic leads}
%\title{Quantum Transport through Resistive  Nanocontacts:\\
%Effective One-Dimensional Theory and Conductance Formulas for Resistive Leads.}
%\title{Quantum Transport through Resistive  Nanocontacts:\\
%Effective One-Dimensional Theory and Conductance Formulas for Non-Ballistic Contacts.}
%\title{Quantum transport through resistive  nanocontacts:\\
%effective one-dimensional theory and generalized Fisher-Lee formula}
%\title{Effective one-dimensional theory and generalized Fisher-Lee formula\\for quantum transport at nanoscale contacts}
%\title{Quantum transport effective one-dimensional theory and generalized Fisher-Lee formula}
%\title{Effective one-dimensional theory of nanocontacts and generalized Fisher-Lee formula}
%\title{Effective one-dimensional theory of quantum transport and generalized Fisher-Lee formula}
%\title{Effective one-dimensional theory of quantum transport at nanoscale contacts: generalized Fisher-Lee formula }

\author{Pierre \surname{Darancet}}
\affiliation{Institut N\'eel, CNRS \& UJF, 38042 Grenoble, France}
\affiliation{European Theoretical Spectroscopy Facility (ETSF)}
\affiliation{Molecular Foundry, Lawrence Berkeley National Laboratory, Berkeley, CA 94720, USA}
\author{Valerio \surname{Olevano}}
\affiliation{Institut N\'eel, CNRS \& UJF, 38042 Grenoble, France}
\affiliation{European Theoretical Spectroscopy Facility (ETSF)}
\author{Didier \surname{Mayou}}
%\email[corresponding author: ]{didier.mayou@grenoble.cnrs.fr}
\affiliation{Institut N\'eel, CNRS \& UJF, 38042 Grenoble, France}
\affiliation{European Theoretical Spectroscopy Facility (ETSF)}

\date{\today}

\begin{abstract}
We introduce a new quantum transport formalism based on a map of a real 3-dimensional lead-conductor-lead system into an \textit{effective 1-dimensional system}.
The resulting effective 1D theory is an in principle exact formalism to calculate the conductance. Besides being more efficient than the principal layers approach, it naturally leads to a \textit{5-partitioned} workbench (instead of 3) 
where each part of the device (the true central device, the ballistic and the non-ballistic leads) is explicitely treated, allowing better physical insight into the contact resistance mechanisms. 
Independently , we derive a \textit{generalized Fisher-Lee formula} and a \textit{generalized Meir-Wingreen formula} for the correlated and uncorrelated conductance and current of the system where the initial restrictions to ballistic leads are generalized to the case of resistive contacts. We present an application to graphene nanoribbons.
\end{abstract}

\maketitle

% Valerio : je propose de ne pas citer les papiers orginaires Kubo, Landauer, Meir & Wingreen, car deja overcites connus par tout le monde, a la limite citer les livres Datta et Di Ventra, ou seulement Di Ventra si ca reprende tout Datta. Kubo Landauer ne vont pas se facher (en tout cas leur formules sont nommees avec leur noms) Datta il est overcite. Utiliser les citations pour les travaux moins standard model (Kurth, Todorov, etc.) mais en tout cas importants, et pour bien disposer les auteurs au cas ou ils etaient referees. Est il le cas de citer Godby?
% Valerio : il faut mettre une footnote avec la question de la convergence aymptotic seulement pour les systemes metallique, sans gap.
% Valerio : il faut mettre une note pour discuter extensions a NEGF, MeirWingreen en presence des correlations dans la molecule centrale. et aussi dans le leads (au moins la renormalisation de niveaux due a la partie hermitique de la self-energie).
% Valerio : discuter extensions a bases non otrthnormees: pas le cas, vue que l'approche se refere a H avec hopping a n'importe quels voisins. Donc en principe on peut utiliser des fonctions de Wannier non necessairement maximally-localised.

\section{Introduction}
%\subsection{General Framework}
The problem to describe nanoscale electronic transport \cite{DiVentra,Datta} from first principles still remains a formidable challenge.
Although powerful formalisms have been developed in the last years \cite{KurthRubioGross,DiVentraTodorov} the cornerstone is still represented by an efficient coupling between electronic structure theories with appropriate modelling of the quantum transport problem. \cite{TaylorGuo,DiVentraPantelides}
In the framework of the Kubo formalism \cite{Datta,DiVentra} very efficient methods, such as the MKRT approach, \cite{Mayou} have allowed a wealth of applications even to technological systems. \cite{Roche}
The Landauer formalism and its NEGF extension \cite{Datta,DiVentra} recently witnessed more intense developments which have led to the setup of a {\it standard model}, often referred as principal layers approach. \cite{Datta}
In this approach, the workbench model is assumed to be a 3-partitioned system \cite{CaroliNozieres} constituted by a central region sandwiched between two semi-infinite leads.
The latter, assumed ballistic and at partial equilibrium, only inject and harvest electrons into the central region where all the processes affecting the conductance are assumed to take place.
Those typically include contact resistance, scattering by impurities and defects, incoherent transport electron-electron and electron-phonon scatterings.
Once provided the electronic structure of the system by an appropriate ({\it ab initio} or semi-empirical) theory, the consecutive quantum transport problem can be solved by e.g. the calculation of the Green's function of the central region in presence of the effect of the leads, represented by the self-energies of semi-infinite periodic systems.
The conductance can then be calculated via the Fisher-Lee or the Meir-Wingreen formulas \cite{Datta} involving the Green's function and the leads injection rates.

%\subsection{Criticisms to the standard model}
One may notice that in this standard model the separation between the central region and the ballistic leads appears somehow arbitrary and unphysical.
Indeed, in order to correctly describe the contact resistance, % that arises at the device-contact interface,
the central region should contain not only the conductor under study (e.g. a molecule or a nanodevice), but also some layers in reality belonging to the leads.
Convergence should be checked by increasing the central region size and thus the number of states in the problem.
Its computational resolution is heavier since it deals with a number of channels much greater than the true channels of the central device.
%Furthermore, the natural physical separation between the leads and the true real device is lost, and a direct comprehension of resistance mechanisms is difficult.

\begin{figure}
 \includegraphics[clip,width=0.48\textwidth]{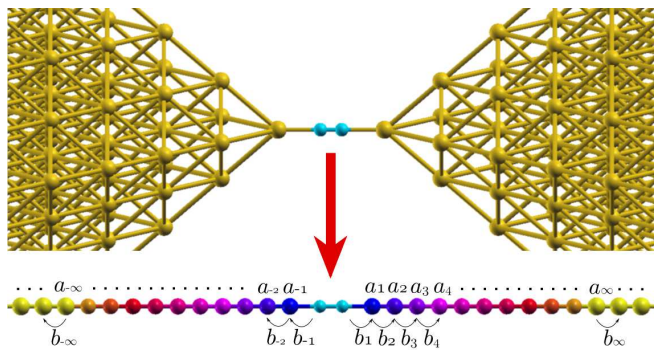}
 \caption{Mapping of a real 3D device (top: hydrogen molecule in between gold leads) into the \textit{effective 1D system} (bottom: effective atomic chain).
 The \textit{effective channels} arise from the central device (here the hydrogen molecule) and pursue into a non-ballistic section (blue, violet and red pseudoatoms), until they achieve an asymptotic ballistic behaviour (yellow).}
 \label{1Deff}
\end{figure}

%\subsection{Effective 1D theory, effective channels, 5-partitioned system, generalized Fisher-Lee formula}
In this work we introduce the notion of \textit{effective channels} as the states through which the current flows up to the central device.
The number of these channels is upper-bounded by the number of states of the central bottleneck.
All the leads' states orthogonal to the states of the effective channels do not effectively participate to the conductance and can be safely disregarded, acting as a prefiltering of the initial problem.
%Pierre - as a prefiltering of the initial problem.
This is a considerable simplification with respect to the principal layers, where the number of the channels is usually much greater than the number of channels in the device.
%Instead, the new approach deals with the minimum number of channels contributing to the current.
The numerically efficiency is consequently much improved.
Moreover, the effective channels can be viewed as associated to an \textit{effective 1-dimensional system} into which the real 3-dimensional physical system is mapped (Fig.~\ref{1Deff}).
This is a way to restore the natural dimensionality of the quantum transport problem, which is truly 1D.
The resulting 1D effective theory is an in principle exact formalism to calculate the conductance.
The main advantages of the 1D approach are:
\begin{itemize}
 \item It is formally exact, and its numerical implementation requires no approximation.
 \item It reduces the size of the numerical problem.
 \item It is particularly efficient for nanocontacts.
 \item The calculation scales as $N_{\text{tot}} \times N_{\text{channels}}^2$, with $N_{\text{tot}}$ the full size of the initial Hamiltonian and $N_{\text{channels}}$ the number of effective channels.
 \item It is independent from the level of theory used to calculate the electronic structure, whether with or without correlation. It relies on the preliminary knowledge of the Hamiltonian and eventual interaction self-energies.
 \item It is independent from the type of basis set used in the previous electronic structure calculation.
\end{itemize}
Indeed, the formalism only requires a complete $N_{\text{channels}}$ set of independent states $\phi_c^i$ localized at the bottleneck, as well as how $H$ acts on states, as implemented in common DFT codes.
Afterwards, there is absolutely no requirement on the basis set that can be atomic orbitals, Wannier functions, non-orthogonal or non-maximally localized, even plane waves.
In fact, another important result of this formalism is that it proposes its own \textit{recursion states} basis as the best suited to solve the quantum transport problem.

Present limitations and drawbacks of the approach are:
\begin{itemize}
 \item It has one step more (the recursion) than a principal layers calculation.
 \item Although a generalization to the $k_\parallel$-dependent transport is possible, the approach is inefficient on non-bottleneck, such as planar, geometries.
 \item For transport at finite voltage, the effective channels must be recalculated at each different bias.
 \end{itemize}

Independently, in this work we will also introduce a physically more intuitive \textit{5-partitioned} (instead of 3-) \textit{new quantum transport workbench} model (Fig.s~\ref{SGFL} and \ref{5PW}).
This is composed by the true central conductor device, the left and right sections of non ballistic leads --- which contain and isolate contact resistance mechanisms --- and finally the ballistic semi-infinite leads.
We will derive the exact uncorrelated and correlated conductance formulas associated to the general workbench where a section of the leads is non-ballistic. 
For the uncorrelated case, we will derive a \textit{generalized Fisher-Lee formula} to be associated with the 5-partitioned workbench.
For the case of correlated transport we will derive a \textit{generalization of the Meir-Wingreen formula} to resistive nanocontacts.
In both cases the effect of the contact resistance is exactly taken into account and contained into \textit{renormalized lead injection rates} $\tilde{\Gamma}$ and \textit{scattering functions} $\tilde{\Sigma}^{<>}$.
The new workbench allows a valuable insight on the origin of resistance into the system and provides a clear analysis of the different resistance mechanisms.
While in the principal layers all the resistance is contained in the extended central region Green's function $G_C$, here
the contact resistance can be directly read from $\tilde{\Gamma}$ and $\tilde{\Sigma}^{<>}$ whereas other more internal  mechanisms --- i.e. density of states effects, scattering by impurities or defects, and e-e and e-ph scattering effects --- can be read from the Green's function $G_c$ of the true central part.
The application of the generalized Fisher-Lee and Meir-Wingreen formulas is not at all restricted to an effective $1$D problem and can be straightforwardly implemented also in a ordinary 3D quantum transport geometry (Fig.~\ref{5PW}).
Any actual principal layer code can be modified in order to refer to the 5-partitioned workbench and the associated quantities.
With respect to the principal layers, the 5-partioned workbench model seems to present only advantages:
\begin{itemize}
\item It allows a clear analysis of the resistance by disentangling contact resistance from other mechanisms.
\item The initial problem is divided into several independent parts, with reduced calculation cost.
\item It is very efficient on the effective 1D system.
\end{itemize}

We will show an application to graphene nanoribbons in the tight-binding approach.

\section{Definition of effective channels}

Let's first provide a heuristic approach to the notion of effective channel.
A generic nanoscale quantum transport system is always characterized by a bottleneck of a few $N_a$ atoms, each contributing with say max $N_o$ orbitals.
For example the bottleneck is represented by a hydrogen atom in Fig.~\ref{1Deff}, a nitrogen atom in Fig.~\ref{5PW} and by 6 carbon atoms in Fig.~\ref{nanoribbon}.
The conductance of the whole system can never exceed $N_{\text{channels}} = N_a N_o$ quantum of conductance.
In the ballistic case the conductance has the characteristic steplike integer profile (Fig.~\ref{nanoribbon}, dashed line), at most summiting to $N_a N_o$.
In the other cases, the contact resistance always reduces the conductance (Fig.~\ref{nanoribbon}, continuous line).
The bottleneck represents an uppermost bound.
So we can say that there are at most $N_a N_o$ channels effectively contributing to the conductance, the rest being idle.
This notion of effective channels can thus be exploited to simplify the problem.

Let's now find a general condition for a state to contribute to transport to be used as definition of effective channel.
We consider a quantum transport system separated into a true central device $c$ (for example ``only'' the hydrogen molecule  of Fig.~\ref{1Deff}) coupled to left $l$ and right $r$ contacts.
The Hamiltonian can be written:
\begin{equation}
 H = H_{c} + H_{l} + H_{r} +  H_{lc} + H_{cl} + H_{cr} + H_{rc}
 ,
 \label{Hamil}
\end{equation}
where $H_{c}$, $H_{l}$ and $H_{r}$ are the Hamiltonians of the central device, the left and right contacts respectively, and $H_{cl} = H_{lc}^\dag$ and $H_{rc} = H_{cr}^\dag$ the coupling of the central device to the left and right contacts.
The absence of direct coupling between the two contacts is here a fundamental requirement.
An electron must pass across the central device to go from one contact to the other.
We name $\mathcal{S}_c$, $\mathcal{S}_l$ and $\mathcal{S}_r$ the Hilbert spaces of states of the central device, left and right contacts.
%We define $U(t)$ as the time evolution operator associated to the full $H$ hamiltonian, while
%$U_{d}(t)$ as the ones associated to the decoupled diagonal Hamiltonian $H_d$.
%Under $U_{d}(t)$ the states evolve separately in three disconnected spaces.
%Let $\phi$ be a state belonging to the left or right contact. If $\phi$ is such that at any time $U_{d}(t)|\phi\rangle$ is {\it not} coupled to the central device, i.e. $\forall t, \hcoupl U_{d}(t)|\phi\rangle = 0$, then this state carries no current through the central device.
%Indeed the full evolution of the state is
%$|\phi(t)\rangle =  U_{d}(t)|\phi\rangle + 1/i\hbar \int _0^t dt' U(t-t')\hcoupl U_{d}(t')|\phi\rangle$,
%so that if $\hcoupl U_{d}(t)|\phi\rangle = 0$ then $|\phi(t)\rangle = U_{d}(t)|\phi\rangle$ and hence the wavepacket will stay forever within its initial contact region.
%Therefore this state will not contribute to the current and can be safely excluded from quantum transport expressions.

%Let us now analyse which are the states such that $\hcoupl U_{d}(t)|\phi\rangle = 0$. We consider the case of the right contact with Hamiltonian $H_{R}$.
We define the \textit{effective channels space} of lead $t$ ($t=l$ or $r$) by
%spanned by the states obtained by starting from any state $\phi_{c}$ of a basis of the central device, coupling it via $H_{tc}$ to the contact and multiply applying the contact Hamiltonian:
\[
 \mathcal{S}^\textit{eff}_{t} = \Span \{ H_t^n H_{tc} | \phi_c \rangle \}  \qquad 
   \forall \phi_c \in \mathcal{S}_c, \forall n \in \mathbb{N}
 .
\]
$\mathcal{S}^\textit{eff}_t \subset \mathcal{S}_{t}$ is a Hilbert subspace of $\mathcal{S}_{t}$ containing only the states coupled to the central device.
On the other hand, its orthogonal complement $\mathcal{S}_t^\bot$ (such that $\mathcal{S}_t = \mathcal{S}^\textit{eff}_t \oplus \mathcal{S}_t^\bot$, direct sum) contains only states which are $H$-disconnected both from $\mathcal{S}_c$ and $\mathcal{S}^\textit{eff}_t$.
%It is easy to see that the (assumed Hermitian) Hamiltonian $H$ cannot couple $S_t^\bot$ neither to the central device $S_c$, nor to the effective channels $S^\textit{eff}_t$ spaces.
The application of $H$ to a state of $\mathcal{S}_t^\bot$ still belongs to $\mathcal{S}_t^\bot$.
This means that an electron from the reservoir $t$ but in a state belonging to $\mathcal{S}_t^\bot$, will stay and evolve in $\mathcal{S}_t^\bot$ without contributing to a current across the central device.
The calculation of the conductance is not affected by $\mathcal{S}_t^\bot$ that can be safely neglected.
Thus the conductance of the real system is that of a simpler effective system living into $\mathcal{S}^\textit{eff}$, defined
\begin{equation}
  \mathcal{S}^\textit{eff} = \mathcal{S}^\textit{eff}_{l} + \mathcal{S}_{c} + \mathcal{S}^\textit{eff}_{r}
 .
 \label{SEFF}
\end{equation}

\section{1D effective system}

The next step is to demonstrate the existence of this effective system by a correct choice of its basis.
In practice we will provide a Gram-Schmidt or, equivalently, a Haydock recursion \cite{Haydock} algorithm to build an orthonormal basis set $\{\psi_n\}$ for $\mathcal{S}^\textit{eff}$. The representation $H^\textit{eff}$ of the original Hamiltonian on $\{\psi_n\}$ will result at the same time.

To start with, let's restrict to the simplest (scalar) case as in Fig.~\ref{1Deff} where the central bottleneck is a single atom with a single orbital, say $\phi_c$.
%The effective space will contain only one effective channel.
%The extension to the general case is straightforward and is detailed in the appendix.
Let's first build the right $\mathcal{S}^\textit{eff}_r$ effective channel space and its basis set $\{\psi_n\}$.
The first element $\psi_1$ of the basis is given by $b_1 | \psi_1 \rangle = H_{rc} | \phi_c \rangle$.
$b_1$ is chosen as normalization factor for $\psi_1$.
Next we calculate $a_1 = \langle \psi_1| H_r | \psi_1\rangle$.
We then calculate the second element by $b_2 |\psi_2\rangle = H_r |\psi_1\rangle - a_1 |\psi_1\rangle$.
$\psi_2$ is orthogonal to $\psi_1$ and normalized by $b_2$.
At the next and all the following steps we iterate the same procedure, $a_n = \langle \psi_n| H_r | \psi_n\rangle$
%$ | b_{n+1}| = || H_r |\psi_n\rangle - a_n |\psi_n\rangle - b^*_n | \psi_{n-1}\rangle || $ 
and $b_{n+1}|\psi_{n+1}\rangle = H_r |\psi_n\rangle - a_n |\psi_n\rangle - b^*_n | \psi_{n-1}\rangle$.
This is an implementation of the standard recursion method:\cite{Haydock}
\begin{equation}
 H_r|\psi_n\rangle = a_n|\psi_n\rangle + b_{n}^*|\psi_{n-1}\rangle+b_{n+1}|\psi_{n+1}\rangle
 . \label{rec}
\end{equation}
In conclusion we end with an orthonormal basis $\{\psi_n\}$ for $\mathcal{S}^\textit{eff}_r$.
With increasing $n$, the state $\psi_n$ is a linear combination of real orbitals belonging to atoms deeper and deeper in the contact.\cite{Haydock}
The recursion can be stopped at an $n=N$ where the coefficients $a_n$, $b_n$ saturate and converge to an asymptotic regime, $a_\infty$, $b_\infty$ (see Fig.~\ref{convergence}).
From this point the leads are consequently ballistic and associated to states achieving a \textit{maximum spread} into the contact region.
\textit{Notice that this practically and numerically recovers the Landauer's paradigm of \emph{reservoir}, \cite{DiVentra} initially formulated as an \emph{ad hoc} hypothesis.}
On this basis set the Hamiltonian is tridiagonal, with onsite $H^\textit{eff}_{nn} = a_n$ and (only) first neighbours hopping coefficients $H^\textit{eff}_{n,n-1} = b_n$.
%$H^\textit{eff}_{n,n-1} = H^{\textit{eff}*}_{n-1,n}$
It can be seen as associated to an effective 1D pseudoatomic chain (Fig.~\ref{1Deff}).
$H^\textit{eff}$ is in fact the same original Hamiltonian but represented on an orthonormal basis where it is tridiagonal.
Notice that the algorithm does not require to store the basis elements $\psi_n$, but just only the $a_n$ and $b_n$ coefficients on the basis (see Appendix \ref{blockspaces}).

The procedure can now be repeated for the left $l$ contact ($\{\psi_n\}$, $n<0$).
And it can be generalized to the matrix $N_a N_o \ne 1$ case (see Appendix \ref{matrec}). 
In this case $a_n$ and $b_n$ are replaced by block matrices $A_n$ and $B_n$ of size $N_a N_o$.
The Hamiltonian still owns a 1D structure but it is now tridiagonal by blocks in the recursion basis:
\begin{equation}
 H^\textit{eff} = H_c + \sum_{n=-N}^N A_n  
       + \sum_{n=-N}^N (B_n + B^{\dag}_n)
 .
 \label{HEFF}
\end{equation}

\begin{figure}
 \includegraphics[clip,width=0.48\textwidth]{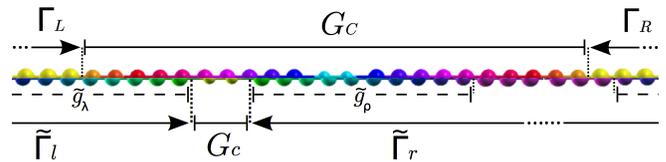}
 \caption{The conductance can be calculated by the traditional Fisher-Lee formula and 3-partitioned workbench, with ballistic leads and extended (molecule + non-ballistic leads) central region (scheme above); or the generalized Fisher-Lee formula and the 5-partitioned workbench, with central device, non ballistic, and finally ballistic sections of the effective channels (scheme below).}
 \label{SGFL}
\end{figure}

\section{Generalized Fisher-Lee formula}

We now focus on the calculation of the conductance of the effective 1D system.
%According to the Landauer viewpoint, when a bias is applied to the system the scattering states coming from one reservoir are occupied up to the reservoir chemical potential. Let us recall that the states that do not transmit across  the central device, i.e. those belonging to $S^{''}_\epsilon$, are occupied up to the chemical potential of the reservoir to which they are connected but do not contribute to the current.
The principal layers approach would reorganize the system into 3 new regions (Fig.~\ref{SGFL} top): a left ballistic lead region L where $a_n$ saturates to the constant asymptotic value $a_{-\infty}$; a right R ballistic lead region $a_{+\infty}$; and an extended central region C containing the true central device $c$ \textit{and also} the two non-ballistic sections of the leads, $n=1,\ldots,N$ and $n=-1,\ldots,-N$.
Referring to this standard workbench, we can calculate the Green's function $g_L$ and $g_R$ of the semi-infinite periodic leads, the associated self-energies $\Sigma_L = H_{CL} g_L H_{LC}$, $\Sigma_R = H_{CR} g_R H_{LR}$ and the injection rates $\Gamma_L$, $\Gamma_R$.
We can then calculate the retarded/advanced Green's function $G^{r/a}_C$ of C in presence of the leads L and R by $G_C(z) = (z - H_C - \Sigma_L - \Sigma_R)^{-1}$.
The conductance is then calculated by the standard Fisher-Lee formula,
\begin{equation}
  C(z) = \frac{2e^2}{h} \tr[ \Gamma_{L}(z) G^{r}_C(z) \Gamma_{R}(z) G^{a}_C(z)] % \frac{2e^2}{h} 
  .
    \label{FL}
\end{equation}
The drawback of this procedure is twofold. On one hand there is no way to analyze separately the role of the true central device and of the non ballistic part of the leads. On the other hand the size of the matrix $G(z)$ can increase rapidly and the matrix inversion can become difficult or impossible, even for a tridiagonal hamiltonian. This restricts the applicability of the method to systems where the resistance is localized at the very vicinity of the nanocontacts.
%These two weaknesses can be circumvented thanks to a new conductivity formula we are going to derive.

\begin{figure}
 \includegraphics[clip,width=0.48\textwidth]{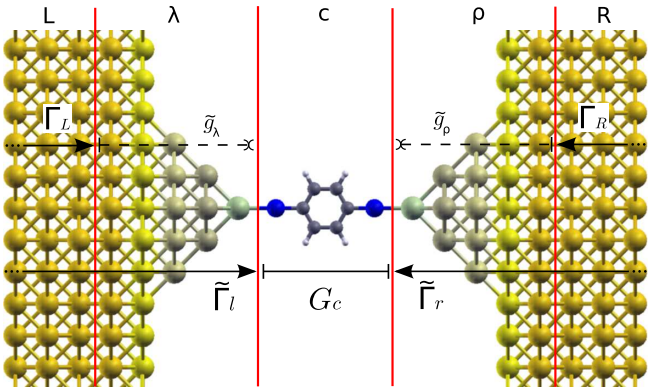}
 \caption{Illustration of the 5-partitioned workbench for a generic 3D quantum transport system, with central device $c$, non ballistic $\lambda, \rho$, and finally ballistic sections L and R of the leads. $\tilde{\Gamma}_{l/r}$ are the contact resistance renormalized lead injection rates, and $\tilde{g}_{\rho/\lambda}$ the Green's functions of the leads non-ballistic sections $\lambda$ and $\rho$ calculated as they were disconnected from the central region c and in presence of the external ballistic leads L and R.}
 \label{5PW}
\end{figure}

In this work we instead propose to keep the original natural separation into true central region $c$ and the leads $l$ and $r$.
Into $l$ and $r$ we identify the ballistic regions $L$ and $R$ and the non-ballistic sections $\lambda$ and $\rho$ (Fig.~\ref{SGFL} bottom and Fig.~\ref{5PW}).
By using the projector $P_c$ on $c$, we single out the Green's function of the true central device $G_c = P_c G_C P_c$.
We then define the Green's function $\tilde{g}_\lambda = [z - P_\lambda (H_C + \Sigma_L) P_\lambda]^{-1}$ (a similar expression for $\tilde{g}_\rho$). This is not the projection of the propagator into the section $\lambda$, $\tilde{g}_\lambda \ne G_\lambda = P_\lambda G_C P_\lambda$. 
Instead it is the propagator in the section $\lambda$ calculated as if $\lambda$ were disconnected from the center $c$ but connected to $L$.
Injecting these definitions into the Fisher-Lee Eq.~(\ref{FL}) and using some fundamental projector relations, \cite{ZM} we get at (see Appendix \ref{gfldemo})
\begin{equation}
  C(z) = \frac{2e^2}{h} \tr [\tilde{\Gamma}_{l}(z) G_{c}^{r}(z) \tilde{\Gamma}_{r} (z) G_{c}^{a}(z)] 
 ,\label{GFL}
\end{equation}
where
\begin{eqnarray}
 \tilde{\Gamma}_{l}(z) &=& H_{cl} \tilde{g}^{a}_{\lambda}(z) \Gamma_{L}(z) \tilde{g}^{r}_{\lambda}(z) H_{lc} ,\nonumber \\
 \tilde{\Gamma}_{r}(z) &=& H_{cr} \tilde{g}^{r}_{\rho}(z) \Gamma_{R}(z) \tilde{g}^{a}_{\rho}(z) H_{rc}
 , \label{gammatilde}
\end{eqnarray}
and
\begin{eqnarray}
 G_c &=& (z - H_c - \tilde{\Sigma}_\lambda - \tilde{\Sigma}_\rho )^{-1} ,\label{Gc}\\
 \tilde{\Sigma}_\lambda &=& H_{cl} \tilde{g}_{\lambda}(z) H_{lc} ,\nonumber \\
 \tilde{\Sigma}_\rho &=& H_{cr} \tilde{g}_{\rho}(z) H_{rc} .\label{Srho}
\end{eqnarray}
Formula (\ref{GFL}) has a Fisher-Lee like form but involves different quantities (for instance $\tilde{\Gamma}_{l/r} \ne i [ \tilde{\Sigma}^r_{\lambda/\rho} - \tilde{\Sigma}^a_{\lambda/\rho}]$).
It now refers to a workbench where the Green's function $G_C$ of the extended central region is replaced by the more significative Green's function $G_c$ of the true device under study.
The injection rates $\Gamma_{L/R}$ of ballistic leads are replaced by \textit{contact resistance dressed renormalized injection rates} $\tilde{\Gamma}_{l/r}$ which refer to both the ballistic L/R and the non ballistic $\lambda$/$\rho$ sections of the leads.
In the principal layers approach all resistance mechanisms are localized within the extended central region C and considered in $G_C$.
Here contact resistance is separated from other mechanisms, localized in the non-ballistic sections $\lambda$ and $\rho$ and transferred into $\tilde{\Gamma}$ where it is taken into account via $\tilde{g}$.
The contact resistance can be read directly from $\tilde{\Gamma}$.
As we will see in the example, if $\tilde{\Gamma}(E)=0$ at a given $E$, this will provide 0 conductance whether or not there is at $E$ an available channel in the central device. Therefore the generalized Fisher-Lee formula allows a more clear interpretation of resistance mechanisms.
Notice that the $\tilde{\Gamma}_{l/r}$ depend only on the electronic structure of the contact and on its coupling to the central device.
Thanks to recurrence relations the calculation of the 1D $\tilde{g}$ and hence of $\tilde{\Gamma}$ can be carried out in a very efficient numerical way (see Appendix \ref{gfldemo}) wrt the principal layers approach where the calculation of $G_C$ can be cumbersome.
Moreover, the generalized Fisher-Lee formula is not restricted to the 1D-effective theory presented here.
It can be applied also to real 3D systems, in place of the ordinary Fisher-Lee, provided the extended central device is split into two non-ballistic leads regions and the true central device (Fig.~\ref{5PW}).
One might think for example about a molecular junction where our formula could be used to project the whole transport problem into molecular orbitals.

\section{Extensions to NEGF and correlated transport}

The 5-partitioned workbench is particularly convenient in the case of correlated transport within NEGF.
Starting from the Meir-Wingreen formula \cite{DiVentra} for the current, the non-coherent term can be separated from the coherent. For the coherent term we end again to the generalized Fisher-Lee Eq.~(\ref{GFL}). For the non-coherent current we can derive a generalized Meir-Wingreen expression of the form:
%($t=l$ or $r$):
\begin{equation}
 i_t^\textrm{noncoh} = \frac{e}{h} \tr [\tilde{\Sigma}_t^< G^r_c \Sigma^>_\textit{corr} G^a_c - \tilde{\Sigma}_t^> G^r_c \Sigma^<_\textit{corr} G^a_c ]
 ,
 \label{GMW}
\end{equation}
where $\Sigma^{<>}_\textit{corr}$ are the in/out scattering functions related only to correlations (e-e or e-ph), while $\tilde{\Sigma}^{<>}_t$ are the in/out \textit{contact resistance renormalized lead $t$ scattering functions}, $\tilde{\Sigma}_t(z) =H_{ct} \tilde{g}^{a}_\tau(z) \Sigma_T(z) \tilde{g}^{r}_\tau(z) H_{tc}$, with $\Sigma_T(z)$ calculated in the ballistic region (see Appendix \ref{gmwdemo}).
Notice that $\tilde{\Sigma}_t \ne \tilde{\Sigma}_\tau$, with $\tilde{\Sigma}_\tau$ defined in Eq.~(\ref{Srho}).
The equilibrium relations $\tilde{\Sigma}^<_t = f^{\rm FD}_T \tilde{\Gamma}_t$ and $\tilde{\Sigma}^>_t = (1 - f^{\rm FD}_T) \tilde{\Gamma}_t$ (as well as $\tilde{\Gamma}_t = [\tilde{\Sigma}^>_t - \tilde{\Sigma}^<_t]$) still hold also for the renormalized quantities.
Contact resistance is now physically separated into $\tilde{\Sigma}_t$ with respect to the resistance raising from e-e and e-ph scattering mechanisms, associated and localized into the true central device and $G_c$.
This more faithfully represents the workbench ideal assumption of lost-of-coherence effects only within the true central device, with leads assumed as everywhere perfectly coherent.

\begin{figure}
  \includegraphics[clip,width=0.40\textwidth]{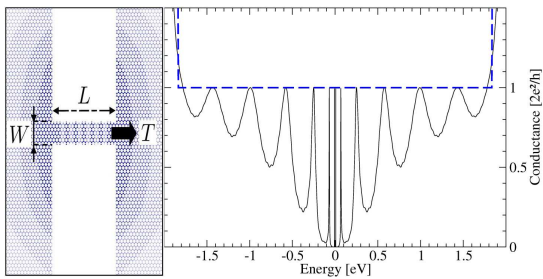}
  \caption{Conductance of a metallic zigzag $<$5.0$>$ graphene nanoribbon (left) of length $L=2.1$ nm connected to graphene 2D seminfinite sheets. The zero of the energy corresponds to the charge neutrality point of graphene i.e. to the Dirac point.}
  \label{nanoribbon}
\end{figure}

\section{Application to a graphene nanodevice}

\begin{figure}
  \includegraphics[clip,width=0.45\textwidth]{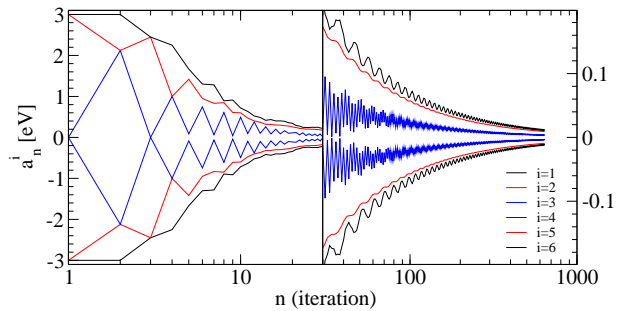}
  \caption{Convergence of the 6 ($i=1,\ldots,6$) $a^i_n$ eigenvalues of the matrix $A$ as function of the iteration $n$.}
  \label{convergence}
\end{figure}

The method has been implemented into a computer code \cite{knesset} and was applied to graphene nanoribbons coupled to graphene sheets \cite{Graphene1,Graphene2} using a tight-binding electronic structure. As shown in Fig.~\ref{nanoribbon} and in Ref.~\cite{Graphene2} the conductance exhibits Fabry-Perot oscillations shorting even to some 0s close to $E=0$. 
Although the nanoribbon is metallic ($A_c(E\simeq 0) \ne 0$) the transmission of electron waves across the contact is blocked at these energies.
It can be shown that the renormalized $\tilde{\Gamma}_{t}(E)$ is zero at those points, whereas $\Gamma_{T}$ keeps finite.
Thank to the new formalism, our analysis shows that the 0 conductance is a pure effect of contact resistance and it can be interpreted as a diffraction effect at the contact constriction (see Ref.~\cite{Graphene2}). 
Notice the high accuracy of the calculation: the maxima of the Fabry-Perot oscillations, in principle exactly equal to $1$, are found equal to $1 \pm 10^{-3} \sim 10^{-4}$. Furthermore, at difference with respect to the principal layers approach as implemented e.g. in WanT \cite{want} or other codes, the conductance as function of the energy does not present spurious structures and spikes shorting to 0.
Notice also the straightforward, well defined and reliable convergence criterion.
In Fig.~\ref{convergence} we show the typical convergent behaviour of the $a_n$ coefficients as function of the iteration $n$ (the 6 eigenvalues of the $A_n$ matrix in the present case).
The code automatically checks the convergence and stops the recursion when $|a_n - a_{n-1}| < \delta E$ and then calculates the conductance.
To get the $10^{-3}$ eV accuracy we needed $n=600$ iterations.
The case studied is however one of the most unfavourable, due to a divergence of the electron wavelength at $E=0$.
We expect a much quicker convergence for less critical systems, where contact resistance is localized closer to the nanocontacts.

\section{Conclusions}

We have introduced a formalism for quantum transport at resistive nanoscale contacts which relies on the introduction of an effective 1D system and its associated 5-partitioned workbench model.
Moreover, we generalize the Fisher-Lee and the Meir-Wingreen formulas to the case of non-ballistic contacts. 
The formalism is physically more intuitive and numerically more efficient than the standard principal layer approach. 
By its versatility and its theoretical generality, our formalism allows to handle unexplored quantum transport problems, like the exact influence of the experimental contact geometry. From a more fundamental point of view, our work offers new perspectives to discuss the existence of Landauer's reservoirs for real systems.

\section{Acknowledgements}
We thank G.-M. Rignanese, X. Blase, Ch. Delerue, J.-Ch. Charlier, P. Bokes, Z. Zanolli, M. Gr\"uning, A. Ferretti, I. Tokatly, C. Verdozzi, S. Kurth, C.-O. Almbladh, R. Godby, M. Di Ventra and J.B. Neaton for useful discussions.
Computer time has been granted by Ciment/Phynum and IDRIS.
We have been supported by the Grenoble Nanoscience Foundation via the RTRA NanoSTAR project.

\appendix

\section{Mapping into the effective 1D system and matrix recursion}

\subsection{Construction of the block spaces}
\label{blockspaces}
Let us use the convention where the recursion into the right terminal $t=r$ is indicated by positive recursion indices $n>0$, while for the left contact $t=l$ by negative indices $n<0$.
Let us define the space $\mathcal{S}^\textit{eff}_{1}$
\[
 \mathcal{S}^\textit{eff}_{1} = \Span \{ H_{rc} | \phi_c \rangle \},  \qquad 
   \forall \phi_c \in \mathcal{S}_c
 ,
\]
(a similar expression holds for $\mathcal{S}^\textit{eff}_{-1} = \Span \{ H_{lc} | \phi^i_c \rangle \}$).
The application of  $H_r$ to $\mathcal{S}^\textit{eff}_{1}$ generates states that can be decomposed on $\mathcal{S}^\textit{eff}_{1}$ itself and on an orthogonal space that we name $\mathcal{S}^\textit{eff}_{2}$. Successive application of $H_r$ to states of $\mathcal{S}^\textit{eff}_{2}$ generates states that can be decomposed on $\mathcal{S}^\textit{eff}_{1}$, $\mathcal{S}^\textit{eff}_{2}$ and a new orthogonal space $\mathcal{S}^\textit{eff}_{3}$. In general, application of $H_r$ to $\mathcal{S}^\textit{eff}_{n}$ generates states that can be decomposed on $\mathcal{S}^\textit{eff}_{n-1}$, $\mathcal{S}^\textit{eff}_{n}$ and a new orthogonal space $\mathcal{S}^\textit{eff}_{n+1}$. Indeed the Hermicity of $H_r$ prevents the coupling between spaces that are not successive order.

The effective spaces for the left and right terminals are hence given by
\[
  \mathcal{S}^\textit{eff}_r = \bigcup_{n>0} \mathcal{S}^\textit{eff}_n , \qquad \qquad
  \mathcal{S}^\textit{eff}_l = \bigcup_{n<0} \mathcal{S}^\textit{eff}_n 
\]

Let us now define the orthogonal projector $P_{n}=P^{\dag}_{n}=P_{n}^{2}$ on the space $\mathcal{S}^\textit{eff}_{n}$. $A_{n}  = P_{n} H P_{n}$ is the restriction of the Hamiltonian $H$ to the space $\mathcal{S}^\textit{eff}_{n}$ and   $B_{n} = P_{n} H P_{n-1}$ represents the coupling from $\mathcal{S}^\textit{eff}_{n-1}$ to $\mathcal{S}^\textit{eff}_{n}$. The coupling from  $\mathcal{S}^\textit{eff}_{n}$ to $\mathcal{S}^\textit{eff}_{n-1}$ is given by $B^\dag_{n}$. 
One can see that
\begin{eqnarray}
  H_r &=&  \sum_{n > 0} A_n + \displaystyle\sum_{n > 1} (B_n + B^\dag_n) \label{REFF1} \\
  H_l &=&  \sum_{n < 0} A_n + \displaystyle\sum_{n < -1} (B_n + B^\dag_n)
\end{eqnarray}
and 
\begin{eqnarray}
  H_{rc} + H_{cr} &=&  B_1 + B^\dag_1 \\
  H_{lc} + H_{cl} &=&  B_{-1} + B^\dag_{-1}
\end{eqnarray}
Therefore the Hamiltonian restricted to the effective space $\mathcal{S}^\textit{eff} = \mathcal{S}^\textit{eff}_{l} + \mathcal{S}_{c} + \mathcal{S}^\textit{eff}_{r}$ can be written:
\begin{equation}
 H^\textit{eff} = H_c + \sum_{n=-N, n\ne 0}^N A_n  
       + \sum_{n=-N, n\ne 0}^N (B_n + B^{\dag}_n)
 .
\end{equation}

\subsection{Matricial recursion}
\label{matrec}
In the general case the subspace spanned by $H_{tc} | \phi_c \rangle$ with $ \phi_c$ any orbital in the central device, is of dimension $N_a N_o$, that is the characteristic dimension associated to the smallest bottleneck in the central device.
In principle, we can start the recursion from the bottleneck itself by finding $N_a N_o$ linearly independent states $\Phi_{c}(i)$ with $i=1, \ldots, N_a N_o$.
In any case, the dimension of $\mathcal{S}^\textit{eff}_{1}$ or $\mathcal{S}^\textit{eff}_{-1}$, as well as all successive spaces $\mathcal{S}^\textit{eff}_n$, is strictly upper-bounded by $N_a N_o$.
The application of the first iteration recursion step leads to $N_a N_o$ linearly independent states, say $\Psi_{1}(i)$ with $i=1, \ldots, N_a N_o$.
The same for all successive iterations, $\Psi_{n}(i)$.
Then $a_{n}$ and $b_{n}$ are replaced by matrices $A_{n}$ and $B_{n}$ of dimension $N_a N_o$. \cite{Darancet}
The recurrence relations have an analogous form.
We consider here the case for the right terminal $r$ and the positive recursion $n>0$.
The projector on the subspace $\mathcal{S}^\textit{eff}_n$ can be written as
\begin{equation}
 P_n =  \sum_{i} |\Psi_{n}(i) \rangle \langle \Psi_{n}(i) | % = |\Psi_{n} \rangle \langle \Psi_{n} | 
 \label{REFF2}
\end{equation}

Then one get from Eq. (\ref{REFF1}) :
\begin{eqnarray}
&& H_r|\Psi_{n}(i)\rangle = \sum_{j} [A_{n}(j,i)|\Psi_{n}(j)\rangle + \nonumber \\
&& B^\dag_{n}(i,j)|\Psi_{n-1}(j)\rangle + B_{n+1}(j,i)|\Psi_{n+1}(j)\rangle] \label{RecVec}
\end{eqnarray}
with
\begin{eqnarray}
&& \langle\Psi_{n}(i)|\Psi_{m}(j)\rangle = \delta_{n,m}\delta_{i,j} \nonumber \\
&& \textrm{except for } n=0 \qquad |\Psi_{0}(i)\rangle = 0
 \label{RecVecexception}
\end{eqnarray}
and
\begin{eqnarray}
 && A_{n}(i,j) =    \langle \Psi_{n}(i)| H |\Psi_{n}(j)\rangle
 \label{REFF3} \\
 && B_{n+1}(i,j) =    \langle \Psi_{n+1}(i)| H |\Psi_{n}(j)\rangle
 \label{REFF4}
\end{eqnarray}
%Defining the vector $\Psi_n = ( \psi_n(i) )$, we can write the fundamental recursion expression Eq.~(\ref{RecVec}) in a matrix form,
%\[
%  H_r|\Psi_{n}\rangle = A_{n}|\Psi_{n}\rangle + B^\dag_{n}|\Psi_{n-1}\rangle + B_{n+1}|\Psi_{n+1}\rangle .\label{RecVecMat}
%\]

The procedure to compute the matrices $A_{n}$ and $B_{n}$ is quite similar to that of the scalar case.  At step $n$ one knows  $A_{m}$ and $B_{m+1}$ for all $m<n$. One stores also the components of $|\Psi_{n}\rangle $ and $|\Psi_{n-1}\rangle $ in the real space basis.

Then $A_{n}(i,j) $ is calculated from  equation (\ref{REFF3}). One then compute the components in the real space basis of $|\Psi^{'}_{n+1}(j)\rangle$:
\begin{eqnarray}
|\Psi^{'}_{n+1}(i)\rangle= \sum_{j} B_{n+1}(j,i)|\Psi_{n+1}(j)\rangle = \nonumber \\
H_{r}|\Psi_{n}(i)\rangle - \sum_{j} [A_{n}(j,i)|\Psi_{n}(j)\rangle - \nonumber \\
B_{n}^\dag(i,j)|\Psi_{n-1}(j)\rangle ]
 \label{REFF5}
\end{eqnarray}
with the overlap matrix
\begin{equation}
S_{n+1}(i,j) =    \langle \Psi^{'}_{n+1}(i) |\Psi^{'}_{n+1}(j)\rangle = (B^\dag_{n+1}B_{n+1})(i,j)
 \label{REFF6}
\end{equation}
Once the overlap matrix  $S_{n+1}(i,j)$ is calculated, we can construct the orthonormal basis of the $\Psi_{n+1}(j)$ states whose components on the basis of the $\Psi^{'}_{n+1}(j)$ states are given, as well as of course on the real space basis.
Once the $\Psi_{n+1}(j)$ states are calculated, $ B_{n+1}(i,j) $ is calculated from Eq.~(\ref{REFF4}).
We then pass to the next step of recursion.

There is a freedom on the choice of the orthonormal  basis of the subspaces $\mathcal{S}^\textit{eff}_{n+1}$ which is generated by $\Psi_{n+1}(j)$ for all $j$, or equivalently by  $\Psi^{'}_{n+1}(j)$.
Indeed the matrices $A_{n}$ and $B_{n}B^{\dag}_{n}$ are defined up to a unitary transformation which depends on the precise choice of the vectors $\Psi_{n}(i)$.
This unitary transformation is equivalent to the freedom in the choice of the phase of the $b_{n}$ coefficients in the scalar case.

The most time consuming part  is the computation of the Hamiltonian  of the effective 1D system.
This amounts essentially to compute $2 N_a N_o$ scalar recursion procedures, where $N_a N_o$ is the dimension of the spaces $\mathcal{S}^\textit{eff}_{n}$ corresponding to the size of the blocks of the tridiagonal Hamiltonian $H^\textit{eff}$.
Since the only operations done on the real space basis are scalar products, the whole procedure exactly scales linearly with the size of the initial Hilbert subspace.
That is, the proposed algorithm is an $O(N)$ method.
Once the Hamiltonian Eq.~(\ref{REFF1}) is computed, the calculation of the operators $\tilde{\Gamma}_t(z)$ and $\tilde{\Sigma}_t(z)$ is relatively fast.

\section{Derivation of the generalized Fisher-Lee and Meir-Wingreen formulas}

In this appendix we present two different schemes of derivation for the generalized Fisher-Lee and Meir-Wingreen formulas.
The first scheme is an ordinary matrix derivation, while the second uses projector techniques.

\subsection{From the 3- to the 5-partitioned workbench}
The original Hamiltonian Eq.(\ref{Hamil}) 
\begin{equation}
 H = 
  \left( \begin{array}{ccc}
   H_l    & H_{lc} & 0      \\
   H_{cl} & H_c    & H_{cr} \\
   0      & H_{rc} & H_r
  \end{array} \right) \label{Hlcr}
\end{equation}
is written in the physically intuitive l--c--r 3-partition of the system, with $c$ the true central region, and $l$ and $r$ the external leads containing both ballistic and non-ballistic sections

In the principal layers approach, the Hamiltonian is divided differently and is rewritten as
\[
 H = 
  \left( \begin{array}{ccc}
   H_L    & H_{LC} & 0      \\
   H_{CL} & H_C    & H_{CR} \\
   0      & H_{RC} & H_R
  \end{array} \right),
\]
that is using a different L--C--R 3-partitioned scheme where the system is divided into left and right ballistic leads L and R, and a central extended region C that now contains the non ballistic sections of the leads.
The associated Green's function is
\[
 G = 
  \left( \begin{array}{ccc}
   G_L    & G_{LC} & G_{LR} \\
   G_{CL} & G_C    & G_{CR} \\
   G_{RL} & G_{RC} & G_R
  \end{array} \right),
\]
and we use the ordinary principal layers formulas to calculate the conductance ($T = L$ or $R$):
\begin{eqnarray}
 && g_T = (z - H_T)^{-1} \nonumber \\
 && \Sigma_T = H_{CT} g_T H_{TC} \nonumber \\
 && G_C = (z - H_C - \Sigma_L - \Sigma_R)^{-1} \label{GC}\\
 && \Gamma_T = i [ \Sigma^r_T - \Sigma^a_T] \nonumber \\
 && C = \frac{2e^2}{h} \tr [\Gamma_L G^r_C \Gamma_R G^a_C ] \nonumber 
\end{eqnarray}

Starting from the principal layers scheme, we now divide C in 3 subsections: the left non-ballistic lead section $\lambda$, a true central region $c$ and a right non-ballistic lead section $\rho$ (see Fig.~\ref{SGFL} and \ref{5PW}).
Referring back to the l--c--r scheme, l is the left lead comprising both the ballistic L section and the non ballistic section $\lambda$, and r is the right lead comprising both the non ballistic section $\rho$ and the ballistic section R.
We end it up with a 5-partitioned system  L--$\lambda$--c--$\rho$--R (see Fig.~\ref{5PW}) and the associated Hamiltonian
\[
 H = 
  \left( \begin{array}{ccccc}
   H_L    & H_{L\lambda} & 0  & 0 & 0    \\
   H_{\lambda L} & H_\lambda    & H_{\lambda c} & 0 & 0\\
   0 & H_{c\lambda} & H_c    & H_{c\rho} & 0 \\
   0  & 0    & H_{\rho c} & H_\rho & H_{\rho R} \\
   0 & 0 & 0     & H_{R\rho} & H_R
  \end{array} \right).
\]
Comparing with the original Hamiltonian Eq.~(\ref{Hlcr}) we see for example that
\[
 H_{cl} = \left( \begin{array}{cc} 0 & H_{c\lambda} \end{array} \right),
\]
and so on, so that in the next we will confuse $H_{ct}$ with $H_{c\tau}$ ($t=l$ or $r$ and $\tau=\lambda$ or $\rho$), whenever it is clear from the context in which space we are working.
In principle we start from the principal layers approch and we will be always working in the space C.

The extended central Green's function $G_C$ is rewritten in its $\lambda$--c--$\rho$ components:
\[
 G_C = \left( 
\begin{array}{ccc}
 G_\lambda        & G_{\lambda c} & G_{\lambda \rho} \\
 G_{c \lambda}    & G_c           & G_{c \rho} \\
 G_{\rho \lambda} & G_{\rho c}    & G_\rho
\end{array}
\right).
\]
The injection rates $\Gamma_L$ and $\Gamma_R$, which live in the C space, have non-zero elements only in the regions $\lambda$ and $\rho$ respectively, and are of the form
\begin{eqnarray}
 \Gamma_L &=& \left( 
\begin{array}{ccc}
 \Gamma_\lambda & 0 & 0 \\
 0        & 0 & 0 \\
 0        & 0 & 0
\end{array}
\right), \label{GammaL}
\\
 \Gamma_R &=& \left( 
\begin{array}{ccc}
 0 & 0 & 0 \\
 0 & 0 & 0 \\
 0 & 0 & \Gamma_\rho
\end{array}
\right), \label{GammaR}
\end{eqnarray}
like also $\Sigma_L$ and $\Sigma_R$.

\subsection{Generalized Fisher-Lee formula}

Starting from the ordinary Fisher-Lee formula referring to the principal layers approach and the L--C--R workbench,
\[
 C = \frac{2 e^2}{h} T =  \frac{2 e^2}{h} \tr[\Gamma_L G^r_C \Gamma_R G^a_C ]
 ,
\]
using Eq.s~(\ref{GammaL}) and (\ref{GammaR}) we rewrite $T$ as
\begin{equation}
 T = \tr[\Gamma_\lambda G^r_{\lambda \rho} \Gamma_\rho G^a_{\rho \lambda} ]
 . \label{TGlr}
\end{equation}
$G_{\lambda\rho}$ and $G_{\rho\lambda}$ are determined expliciting Eq.~(\ref{GC}) on the $\lambda$--c--$\rho$ scheme,
\begin{eqnarray*}
 && \left[ z - \left(
 \begin{array}{ccc}
 H_\lambda + \Sigma_\lambda & H_{\lambda c} & 0 \\
 H_{c\lambda}               & H_c           & H_{c \rho} \\
 0                          & H_{\rho c}    & H_\rho + \Sigma_\rho
\end{array}
 \right) \right] \cdot \\ && \cdot
 \left( 
\begin{array}{ccc}
 G_\lambda        & G_{\lambda c} & G_{\lambda \rho} \\
 G_{c \lambda}    & G_c           & G_{c \rho} \\
 G_{\rho \lambda} & G_{\rho c}    & G_\rho
\end{array}
\right)
= 1
 .
\end{eqnarray*}

From
\begin{eqnarray*}
 && (z - H_\lambda - \Sigma_\lambda) G_{\lambda c} - H_{\lambda c} G_c = 0 \\
 && -H_{\rho c} G_c + (z - H_\rho - \Sigma_\rho) G_{\rho c} = 0 \\
 && - H_{c \lambda} G_{\lambda c} + (z - H_c) G_c - H_{c\rho} G_{\rho c} = 1
 ,
\end{eqnarray*}
we get
\begin{eqnarray*}
 G_{\lambda c} &=& \tilde{g}_\lambda H_{\lambda c} G_c \\
 G_{\rho c} &=& \tilde{g}_\rho H_{\rho c} G_c \\
 G_c &=& (z - H_c - \tilde{\Sigma}_\lambda - \tilde{\Sigma}_\rho )^{-1}
 ,
\end{eqnarray*}
where we have defined
\begin{eqnarray}
 \tilde{g}_\lambda &=& (z - H_\lambda - \Sigma_\lambda)^{-1} \nonumber \\
 \tilde{g}_\rho &=& (z - H_\rho - \Sigma_\rho)^{-1} \nonumber \\
 \tilde{\Sigma}_\lambda &=& H_{c \lambda} \tilde{g}_\lambda H_{\lambda c} \nonumber \\
 \tilde{\Sigma}_\rho &=& H_{c \rho} \tilde{g}_\rho H_{\rho c} 
 \label{Sigmarho}
 .
\end{eqnarray}
Notice that $\tilde{g}_\tau \ne G_\tau$ (where $\tau = \lambda$ or $\rho$).
The $\tilde{g}_\tau$ can be physically interpreted as the Green's function of the non-ballistic lead section $\tau$ (see Fig.~\ref{5PW}) as disconnected from the central region $c$ ($H_{c\tau} = H_{\tau c} = 0$) but in contact to the ballistic lead region T and under the effect of its $\Sigma_T$ ($\Sigma_\tau$).

Now from
\begin{eqnarray*}
 && (z - H_\lambda - \Sigma_\lambda) G_{\lambda \rho} - H_{\lambda c} G_{c\rho} = 0 \\
 && - H_{\rho c} G_{c\lambda} + (z - H_\rho - \Sigma_\rho ) G_{\rho\lambda} = 0 \\
 && (z - H_\lambda - \Sigma_\lambda) G_\lambda - H_{\lambda c} G_{c \lambda} = 1 \\
 && - H_{\rho c} G_{c\rho} + (z - H_\rho - \Sigma_\rho ) G_\rho = 1
 ,
\end{eqnarray*}
we get
\begin{eqnarray}
 G_{\lambda\rho} &=& \tilde{g}_\lambda H_{\lambda c} G_{c\rho} \label{Glr} \\
 G_{\rho \lambda} &=& \tilde{g}_\rho H_{\rho c} G_{c\lambda} \label{Grl} \\
 G_\lambda &=& \tilde{g}_\lambda +  \tilde{g}_\lambda H_{\lambda c} G_{c\lambda} \nonumber \\
 G_\rho &=& \tilde{g}_\rho +  \tilde{g}_\rho H_{\rho c} G_{c\rho} . \nonumber
\end{eqnarray}
And finally from
\begin{eqnarray*}
 && - H_{c \lambda} G_\lambda + (z - H_c) G_{c\lambda} - H_{c\rho} G_{\rho \lambda} = 0 \\
 && - H_{c \lambda} G_{\lambda \rho} + (z - H_c) G_{c\rho} - H_{c\rho} G_\rho = 0
 ,
\end{eqnarray*}
we get
\begin{eqnarray*}
 G_{c\lambda} &=& G_c H_{c\lambda} \tilde{g}_\lambda \\
 G_{c\rho} &=& G_c H_{c\rho} \tilde{g}_\rho
 ,
\end{eqnarray*}
which replaced into Eq.s~(\ref{Glr}) and (\ref{Grl}), provide
\begin{eqnarray*}
 G_{\lambda\rho} &=& \tilde{g}_\lambda H_{\lambda c} G_c H_{c\rho} \tilde{g}_\rho \\
 G_{\rho\lambda} &=& \tilde{g}_\rho H_{\rho c} G_c H_{c \lambda} \tilde{g}_\lambda
 ,
\end{eqnarray*}
that we can finally insert into Eq.~(\ref{TGlr}) to get the transmittance
\begin{eqnarray*}
 T &=& \tr[\Gamma_\lambda \tilde{g}^r_\lambda H_{\lambda c} G^r_c H_{c\rho} \tilde{g}^r_\rho \Gamma_\rho \tilde{g}^a_\rho H_{\rho c} G^a_c H_{c \lambda} \tilde{g}^a_\lambda ] \\
  &=& \tr[H_{c \lambda} \tilde{g}^a_\lambda  \Gamma_\lambda \tilde{g}^r_\lambda H_{\lambda c} G^r_c H_{c\rho} \tilde{g}^r_\rho \Gamma_\rho \tilde{g}^a_\rho H_{\rho c} G^a_c ]
 .
\end{eqnarray*}
Defining the contact resistance dressed renormalized lead injection rates $\tilde{\Gamma}$,
\begin{eqnarray}
  \tilde{\Gamma}_l &=& H_{c\lambda} \tilde{g}^a_\lambda \Gamma_L \tilde{g}^r_\lambda H_{\lambda c} , \\
 \tilde{\Gamma}_r &=& H_{c\rho} \tilde{g}^r_\rho \Gamma_R \tilde{g}^a_\rho H_{\rho c}
 \label{tildeGamma}
,
\end{eqnarray}
we get the final result for the generalized Fisher-Lee formula
\[
 T = \tr [ \tilde{\Gamma}_l G^r_c \tilde{\Gamma}_r G^a_c ]
 .
\]

Finally we notice that
\[
  \tilde{\Gamma}_{t}  \ne  i[ \tilde{\Sigma}^r_{\tau} - \tilde{\Sigma}^a_{\tau}],
\]
as it can be checked from the definitions Eqs.~(\ref{tildeGamma}) and (\ref{Sigmarho}).
We can however define a new $\tilde{\Sigma}_t \ne \tilde{\Sigma}_{\tau}$,
\[
 \tilde{\Sigma}_t = H_{c\tau} \tilde{g}^a_\tau \Sigma_T \tilde{g}^r_\tau H_{\tau c}
\]
such that the relation
\[
  \tilde{\Gamma}_{t}  =  i[ \tilde{\Sigma}^r_t - \tilde{\Sigma}^a_t],
\]
holds.

\subsection{Fundamental relations for the resolvant}
We now present the projector methodology developped by Zwanzig and Mori \cite{ZM} that we will use later to provide a different derivation of both the generalized Fisher-Lee and Meir-Wingreen formulas.
We will first rederive some fundamental relations that we will use in the next sections.
Let's consider a space $\mathcal{P}$ and its associated projector $P$.
Let's define also its complementary $Q$, that is the projector associated to the complementary space $\mathcal{Q}$.
By their definition, the following relations hold for $P$ and $Q$:
\begin{eqnarray*}
 && P+Q = 1 , \\
 && P^2 = P , \\
 && Q^2 = Q , \\
 && PQ=QP=0
 .
\end{eqnarray*}
We now consider a not necessarily Hermitian operator $H$ (which can even depend on $z$ although we will drop in the following this dependence) and its associated Green's function $G(z) = (z-H)^{-1}$.
Starting from the identity
\[
  (z - H) G = 1
 ,
\]
and multiplying from the left by $P$ and from the right by $Q$, we get the relation
\[
 P (z - H) G Q = PQ = 0
 .
\]
We now insert the identity operator $(P+Q)$,
\begin{eqnarray*}
 0 &=& P (z - H) (P+Q) G Q \\
   &=& P (z-H) PGQ + P (z-H) QGQ \\
   &=& P (z-H) PPGQ - PHQGQ
 ,
\end{eqnarray*}
and we finally arrive to the fundamental projector relation
\begin{equation}
 PGQ = [P(z-H)P]^{-1} PHQGQ
 . \label{zwanzwigmori1}
\end{equation}

Starting from the identity
\[
 Q G (z-H) P = 0
 ,
\]
we arrive to the other Zwanzig-Mori projector relation
\begin{equation}
 QGP = QGQHP[P(z-H)P]^{-1}
 . \label{zwanzwigmori2}
\end{equation}

\subsection{Generalized Fisher-Lee formula by the projector scheme}
\label{gfldemo}

With respect to the $\lambda$--c--$\rho$ regions scheme, we define the projectors $P_\lambda$, $P_c$ and $P_\rho$ which undergo the following relations
\begin{eqnarray}
  && P_\lambda + P_c + P_\rho = 1 , \\ 
  && Q_\lambda = P_c + P_\rho  , \label{ql} \\
  && Q_\rho = P_\lambda + P_c  , \label{qr} \\
  && Q_\lambda P_\rho = P_\rho  , \label{qlpr} \\
  && Q_\rho P_\lambda = P_\lambda  \label{qrpl}
  .
\end{eqnarray}

For a tridiagonal Hamiltonian $H'_C$ (here we use the convention to include into $H'_C$ not only the Hamiltonian $H_C$ associated to the region $C$, but also the lead L and R self-energies, $H'_C = H_C + \Sigma_L + \Sigma_R$), the only non-zero crossed terms are
\begin{eqnarray}
  && H_{\lambda c} = P_\lambda H'_C P_c , \label{hlc} \\
  && H_{c \lambda} = P_c H'_C P_\lambda  , \label{hcl}\\
  && H_{c \rho} = P_c H'_C P_\rho  , \label{hcr}\\
  && H_{\rho c} = P_\rho H'_C P_c  ,\label{hrc}
\end{eqnarray}
while the only non-zero $\Gamma$ elements are:
\begin{eqnarray}
  \Gamma_L &=& P_\lambda \Gamma_L P_\lambda = \Gamma_\lambda
    , \label{plglpl} \\
  \Gamma_R &=& P_\rho \Gamma_R P_\rho = \Gamma_\rho
    . \label{prgrpr} 
\end{eqnarray}

We now start from the ordinary Fisher-Lee formula referring to the principal layers approach and the L--C--R workbench,
\[
 C = \frac{2 e^2}{h} T =  \frac{2 e^2}{h} \tr[\Gamma_L G^r_C \Gamma_R G^a_C ]
 .
\]
We rewrite the transmittance $T$ using Eqs.~(\ref{plglpl}) and (\ref{prgrpr})
\[
 T = \tr [ P_\lambda \Gamma_L P_\lambda G^r_C P_\rho \Gamma_R P_\rho G^a_C ]
 .
\]
Using the cycling property of the trace,
\[
 T = \tr [ \Gamma_L P_\lambda G^r_C P_\rho \Gamma_R P_\rho G^a_C P_\lambda ]
 ,
\]
and the projector properties Eqs.~(\ref{qlpr}) and (\ref{qrpl}), we arrive to
\[
 T = \tr [ \Gamma_L \{ P_\lambda G^r_C Q_\lambda \} P_\rho \Gamma_R \{ P_\rho G^a_C Q_\rho \} P_\lambda]
 .
\]
Now we use the Zwanzwig-Mori relation Eq.~(\ref{zwanzwigmori1})
\begin{eqnarray}
 T &=& \tr \big[ \Gamma_L \left\{ [P_\lambda (z-H'_C) P_\lambda]^{-1} P_\lambda H'_C Q_\lambda G^r_C Q_\lambda \right\} P_\rho \Gamma_R 
   \nonumber \\
   && \left\{ [P_\rho (z-H'_C) P_\rho]^{-1} P_\rho H'_C Q_\rho G^a_C Q_\rho \right\} P_\lambda \big]
 \label{T14} .
\end{eqnarray}
We now define the Green's functions $\tilde{g}$,
\begin{eqnarray}
 && \tilde{g}_\lambda = [P_\lambda (z-H'_C) P_\lambda]^{-1} , \label{gtilde} \label{glambda} \\
 && \tilde{g}_\rho = [P_\rho (z-H'_C) P_\rho]^{-1} \label{grho} .
\end{eqnarray}
Notice that $\tilde{g}_\lambda \ne P_\lambda (z - H'_C)^{-1} P_\lambda$ and $\tilde{g}_\rho \ne P_\rho (z - H'_C)^{-1} P_\rho$.
The $\tilde{g}_\lambda$ ($\tilde{g}_\rho$) can be physically interpreted as the Green's function of the non-ballistic lead section $\lambda$ ($\rho$) (see Fig.~\ref{5PW}) as disconnected from the central region $c$ but in contact to the ballistic lead region L (R) and under the effect of its $\Sigma_L$ ($\Sigma_R$).
Replacing the definition Eqs.~(\ref{glambda}) and (\ref{grho}) into Eq.~(\ref{T14}), we get at
\[
 T = \tr [ \Gamma_L \tilde{g}^r_\lambda P_\lambda H'_C Q_\lambda G^r_C Q_\lambda P_\rho \Gamma_R 
              \tilde{g}^a_\rho P_\rho H'_C Q_\rho G^a_C Q_\rho P_\lambda]
 .
\]
Developping $Q_\lambda$ and $Q_\rho$ according to Eqs.~(\ref{ql}) and (\ref{qr}) and using Eqs.~(\ref{hlc}) and (\ref{hrc}) for $H_{\lambda c}$ and $H_{\rho c}$, we get
\[
 T = \tr [ \Gamma_L \tilde{g}^r_\lambda H_{\lambda c} P_c G^r_C P_\rho \Gamma_R 
              \tilde{g}^a_\rho H_{\rho c} P_c G^a_C P_\lambda]
 .
\]
Using the projector relations $P_c = P_c Q_\rho$ and $P_c = P_c Q_\lambda$,
\[
 T = \tr [ \Gamma_L \tilde{g}^r_\lambda H_{\lambda c} P_c \{ Q_\rho G^r_C P_\rho \} \Gamma_R 
              \tilde{g}^a_\rho H_{\rho c} P_c \{ Q_\lambda G^a_C P_\lambda \}]
 ,
\]
and the Zwanzwig-Mori relation Eq.~(\ref{zwanzwigmori2}), we get
\begin{eqnarray*}
 T  &=& \tr \big[ \Gamma_L \tilde{g}^r_\lambda H_{\lambda c} P_c \left\{ Q_\rho G^r_C Q_\rho H'_C P_\rho [P_\rho(z-H'_C)P_\rho]^{-1} \right\} \\
 && \Gamma_R \tilde{g}^a_\rho H_{\rho c} P_c \left\{ Q_\lambda G^a_C Q_\lambda H'_C P_\lambda [P_\lambda(z-H'_C)P_\lambda]^{-1} \right\} \big] .
\end{eqnarray*}
Using again the definition the Green's functions $\tilde{g}$ and the project relations,
\begin{eqnarray*}
 T &=& \tr [ \Gamma_L \tilde{g}^r_\lambda H_{\lambda c} P_c G^r_C P_c H'_C P_\rho \tilde{g}^r_\rho \\
  &&         \Gamma_R \tilde{g}^a_\rho H_{\rho c} P_c G^a_C P_c H'_C P_\lambda \tilde{g}^a_\lambda]
 ,
\end{eqnarray*}
and the definitions of $H_{c \rho}$ and $H_{c \lambda}$ Eqs.~(\ref{hcl}) and (\ref{hcr}), we get
\[
 T = \tr [ \Gamma_L \tilde{g}^r_\lambda H_{\lambda c} P_c G^r_C P_c H_{c \rho} \tilde{g}^r_\rho \Gamma_R 
              \tilde{g}^a_\rho H_{\rho c} P_c G^a_C P_c H_{c \lambda} \tilde{g}^a_\lambda]
 .
\]
We now define the projection $G_c$ of $G_C$ into c,
\begin{eqnarray}
G_c &=& P_c G_C P_c = P_c (z - H'_C)^{-1} P_c
 , \label{gc} \\
 &=& P_c (z - H_c - \tilde{\Sigma}_{\lambda} - \tilde{\Sigma}_{\rho}  )^{-1} P_c, 
\end{eqnarray}
where
\begin{eqnarray}
 H_c &=& P_c H'_C P_c  , \\
\tilde{\Sigma}_{\lambda} &=& H_{c \lambda} \tilde{g}_\lambda H_{\lambda c}   ,  \\
 \tilde{\Sigma}_{\rho} &=& H_{c \rho} \tilde{g}_\rho  H_{\rho c}, 
\end{eqnarray}
so to get
\[
 T = \tr [ \Gamma_L \tilde{g}^r_\lambda H_{\lambda c} G^r_c H_{c \rho} \tilde{g}^r_\rho \Gamma_R 
              \tilde{g}^a_\rho H_{\rho c} G^a_c H_{c \lambda} \tilde{g}^a_\lambda ]
 .
\]
With a final cycle of the trace,
\[
 T = \tr [ H_{c \lambda} \tilde{g}^a_\lambda \Gamma_L \tilde{g}^r_\lambda H_{\lambda c} G^r_c H_{c \rho} \tilde{g}^r_\rho \Gamma_R 
              \tilde{g}^a_\rho H_{\rho c} G^a_c ] 
 ,
\]
and defining the contact resistance dressed renormalized lead injection rates $\tilde{\Gamma}$,
\begin{eqnarray}
  \tilde{\Gamma}_l &=& H_{c \lambda} \tilde{g}^a_\lambda \Gamma_L \tilde{g}^r_\lambda H_{\lambda c} , \\
 \tilde{\Gamma}_r &=& H_{c \rho} \tilde{g}^r_\rho \Gamma_R \tilde{g}^a_\rho H_{\rho c}
,
\end{eqnarray}
where the generalized $\tilde{\Gamma}_{l/r}$ contain the ballistic leads injection rates $\Gamma_{L/R}$ plus some other ingredients physically containing contact resistance.
We finally get the generalized Fisher-Lee formula
\[
 T = \tr [ \tilde{\Gamma}_l G^r_c \tilde{\Gamma}_r G^a_c ]
 .
\]

\subsection{NEGF fundamental relations}
\noindent

We use Kadanoff and Baym \cite{KadanoffBaym} notations where correlation and scattering functions $G^{<>}, \Sigma^{<>}$ are defined Hermitean.
%We use the same notations as in Datta (\cite{Datta}) except that the symbol $<$ replaces $in$ or $n$ and the symbol $>$ replaces $out$ or $p$.
Let's recall the fundamental relations
\begin{eqnarray}
 \Gamma &=& i(\Sigma^r - \Sigma^a) = \Sigma^< + \Sigma^>
 , \label{A5-1} \\
 A &=& i( G^r -  G^a)= G^< + G^>
 . \label{A5-2}
\end{eqnarray}
In steady-state NEGF, we have two more dynamical equations to get $G^<$ and $G^>$:
\begin{eqnarray}
 G^< &=& G^r \Sigma^< G^a
 , \label{A5-3} \\
 G^> &=& G^r \Sigma^> G^a
 , \label{A5-4}
\end{eqnarray}
where $\Sigma$ is 
\begin{equation}
 \Sigma = \Sigma_{corr} + \sum_T \Sigma_T
 ,
\end{equation}
a sum over the leads $T$ ($T = L$ or $R$ in case of two terminals) and the correlation $\Sigma$.
$\Sigma^{r/a/</>}_{corr}$ is the correlation retarded/advanced self-energy or the in/out scattering function. For ballistic leads $T$ at equilibrium with their reservoirs one has
\begin{eqnarray}
 \Sigma^<_T(E) &=& f^\textrm{FD}_T(E) \Gamma_T(E)
 , \label{A5-7} \\
 \Sigma^>_T(E) &=& (1-f^\textrm{FD}_T(E)) \Gamma_T(E)
 , \label{A5-8}
\end{eqnarray}
where $ \Gamma_T$ is the injection rate of the considered ballistic lead $T$. $f^\textrm{FD}_T(E)$ is the Fermi-Dirac distribution of the reservoir connected to the ballistic lead under consideration.
We also remind the relation
\begin{equation}
A = G^r \Gamma G^a = G^a \Gamma G^r
. \label{A5-5}
\end{equation}

\subsection{Generalized Meir-Wingreen formula}
\label{gmwdemo}

The correlation self-energy acts only in the \textit{true} central device $c$, since one assumes that there are no interactions in the leads (ballistic and non ballistic parts).
Therefore we have the following identity
\begin{equation}
 P_c \Sigma_{corr} P_c = \Sigma_{corr}
 , \label{pcorr}
\end{equation}
while, as before, for the considered terminal T = L or R; $\tau = \rho$ or $\lambda$ we have
\begin{equation}
 P_\tau \Sigma_T P_\tau = \Sigma_T 
 . \label{pst}
\end{equation}
We introduce again the projectors
\[
P_\tau + P_c + P_{\tau'} = 1
 ,
\]
where $P_c$ projects on the \textit{true} central device $c$ and $P_\tau$ and $P_{\tau'}$ project on the non ballistic part of the leads.
We also introduce the conjugated projector
\[
Q_\tau  = 1 - P_\tau
.
\]

The Meir-Wingreen formula for the current $ i_T dE$ that enters in the central device through the terminal $T$ per energy interval $dE$, is:
\begin{eqnarray*}
 i_T &=& \frac{e}{h} \tr [\Sigma^<_T G^>_C - \Sigma^>_T G^<_C] \\
  &=& \frac{e}{h}  \tr [\Sigma^<_T G^r_C \Sigma^> G^a_C - \Sigma^>_T G^r_C \Sigma^< G^a_C]
 .
\end{eqnarray*}
$i_T$  can be divided into the coherent and non-coherent parts,
\begin{equation}
 i_T = i_T^{\rm coh} + i_T^{\rm ncoh}
 ,
\end{equation}
where the non-coherent part of the Meir-Wingreen formula is defined
\begin{equation}
 i_T^{\rm ncoh} = \frac{e}{h} \tr [\Sigma^<_T G^r_C \Sigma^>_{corr} G^a_C - \Sigma^>_T G^r_C \Sigma^<_{corr} G^a_C ]
 , \label{incoh}
\end{equation}
while, using the fundamental relations of NEGF, the coherent part can be written
\begin{equation}
 i_T^{\rm coh} = \frac{e}{h} \tr [\Gamma_T G^r_C \Gamma_{T'} G^a_C ] (f_T^{\rm FD} - f_{T'}^{\rm FD})
 ,
\end{equation}
for the specific case of two terminals $T$ and $T'$.
$f_T^{\rm FD}$ is the Fermi-Dirac distribution of lead $T$ (always at equilibrium).
The latter can be recognized as the ordinary Fisher-Lee formula when associating $T=L$ and $T'=R$,
\begin{equation}
 i^{\rm coh} = \frac{e}{h} \tr [\Gamma_L G^r_C \Gamma_R G^a_C ] (f_L^{\rm FD} - f_R^{\rm FD})
 .
\end{equation}

To generalize the Meir-Wingreen formula we refer again to the the l--c--r scheme and the associated quantities, the Green's function of the $\it {true}$ central device and the renormalized injection rates of the leads.
The derivation for the generalized Meir-Wingreen formula proceeds separately on the coherent and the non-coherent part.
For the coherent part, the derivation gets back to the generalized Fisher-Lee formula:
\begin{equation}
 i_T^{coh} = \frac{e}{h} \tr [\tilde{\Gamma}_t G^r_c \tilde{\Gamma}_{t'} G^a_c ] (f^{\rm FD}_T - f^{\rm FD}_{T'})
 .
\end{equation}

For the non coherent part, we start from Eq.~(\ref{incoh}).
Let's for the moment consider only the first term of Eq.~(\ref{incoh}),
\[
 T_1 = \tr [\Sigma^<_T G^r_C \Sigma^>_{corr} G^a_C ]
 .
\]
Using Eqs.~(\ref{pcorr}) and (\ref{pst}), we get
\[
 T_1 = \tr [P_\tau \Sigma^<_T P_\tau G^r_C P_c \Sigma^>_{corr} P_c G^a_C]
 ,
\]
and cycling the trace and using projector properties
\[
 T_1 = \tr [\Sigma^<_T \{ P_\tau G^r_C Q_\tau \} P_c \Sigma^>_{corr} P_c \{ Q_\tau G^a_C P_\tau \} ]
 .
\]
We now use the Zanzwig-Mori relations Eqs.~(\ref{zwanzwigmori1}) and (\ref{zwanzwigmori2}), and the definition of the $\tilde{g}_\tau$ Green's function Eq.~(\ref{gtilde}), so to get
\[
 T_1 = \tr [\Sigma^<_T \tilde{g}^r_\tau P_\tau H_C Q_\tau G^r_C Q_\tau P_c \Sigma^>_{corr} P_c Q_\tau G^a_C Q_\tau H_C P_\tau \tilde{g}^a_\tau ]
 .
\]

Now using the definition of $H_{\tau c}$ and $H_{c\tau}$ Eqs.~(\ref{hlc}) and (\ref{hcl}),
\[
 T_1 = \tr [\Sigma^<_T \tilde{g}^r_\tau H_{\tau c} P_c G^r_C P_c \Sigma^>_{corr} P_c G^a_C P_c H_{c \tau} \tilde{g}^a_\tau ]
 ,
\]
and the definition of $G_c$ Eq.~(\ref{gc}),
\[
 T_1 = \tr [\Sigma^<_T \tilde{g}^r_\tau H_{\tau c} G^r_c \Sigma^>_{corr} G^a_c H_{c \tau} \tilde{g}^a_\tau ]
 ,
\]
we get, after cycling of the trace, at
\[
 T_1 = \tr [ H_{c \tau} \tilde{g}^a_\tau \Sigma^<_T \tilde{g}^r_\tau H_{\tau c} G^r_c \Sigma^>_{corr} G^a_c ]
 .
\]
We now define the renormalized lead injection rates,
\begin{equation}
 \tilde{\Sigma}_t = H_{c \tau} \tilde{g}^a_\tau \Sigma_T \tilde{g}^r_\tau H_{\tau c}
 ,
\end{equation}
so that $T_1$ reduces to
\[
 T_1 = \tr [ \tilde{\Sigma}^<_t G^r_c \Sigma^>_{corr} G^a_c ]
 .
\]
When considering also the second term in Eq.~(\ref{incoh}), we get finally at our generalized Meir-Wingreen formula,
\[
 i_T^{\rm ncoh} = \frac{e}{h}  \tr [ \tilde{\Sigma}^<_t G^r_c \Sigma^>_{corr} G^a_c - \tilde{\Sigma}^>_t G^r_c \Sigma^<_{corr} G^a_c ]
 .
\]

\end{document}